# Multiferroic tunnel junctions

Martin Gajek[1,2], Manuel Bibes[3], Stéphane Fusil[1], Karim Bouzehouane[1], Josep Fontcuberta[2], Agnès Barthélémy[1] & Albert Fert[1]

[1] Unité Mixte de Physique CNRS-Thales, Route départementale 128, 91767 Palaiseau, France

[2] Institut de Ciència de Materials de Barcelona, CSIC, Campus de la Universitat Autonoma de Barcelona, 08193 Bellaterra, Spain

[3] Institut d'Electronique Fondamentale, Université Paris-Sud, 91405 Orsay, France

**Multiferroics are singular materials that can display simultaneously electric and magnetic orders [1,2]. Some of them can be ferroelectric and ferromagnetic and, for example, provide the unique opportunity of encoding information independently in electric polarization and magnetization to obtain four different logic states. However, schemes allowing a simple electrical readout of these different states have not been demonstrated so far. In this article, we show that this can be achieved if a multiferroic material is used as the tunnel barrier in a magnetic tunnel junction. We demonstrate that thin films of ferromagnetic-ferroelectric $La_{0.1}Bi_{0.9}MnO_3$ (LBMO) retain both ferroic properties down to a thickness of only 2 nm. We have used such films as spin-filtering tunnel barriers the magnetization and electric polarization of which can be switched independently. In that case, the tunnel current across the structure is controlled by both the magnetic and ferroelectric configuration of the barrier, which gives rise to four distinct resistance states. This can be explained by the combination of spin filtering by the ferromagnetic LBMO barrier and the partial charge screening of electrical charges at the barrier/electrode interfaces due to ferroelectricity. We anticipate our results to be a starting point for more studies on the interplay**

**between ferroelectricity and spin-dependent tunneling, and for the use of nanometric multiferroic elements in prototype devices. On a wider perspective, they may open the way towards novel reconfigurable logic spintronics architectures and to electrically controlled readout in quantum computing schemes using the spin-filter effect [3].**

The research on magnetic multilayers in the 1980's [4] has led to a new type of electronics exploiting the spin of the carriers, the so-called spintronics [5]. Simultaneously, advances in ferroelectric thin film research have led to several technological applications in sensor industry and consumer electronics [6]. While magnetism and ferroelectricity usually exclude each other [1], it is known since the late 1960's that they can indeed coexist in a few materials called multiferroics [2]. These compounds exhibit magnetic and electric orders and thus provide a unique opportunity to exploit several functionalities in a single material [7,8].

A way to simply exploit this multifunctional character, which as never been reported yet, is to design magnetic tunnel junctions integrating a nanometric ferromagnetic-ferroelectric film as the tunnel barrier, a key prerequisite being however the stability of ferroelectricity at the very small thickness of a tunnel barrier [9].

Here, we report on the fabrication and characterization of magnetic tunnel junctions including a 2 nm-thick barrier made of the ferromagnetic and ferroelectric material $La_{0.1}Bi_{0.9}MnO_3$ (LBMO). These tunnel junctions allow us to implement simultaneously two recently proposed concepts, namely spin-filtering by a ferromagnetic barrier (the so-called spin-filter effect [10,11]), and the influence of the ferroelectricity on the tunneling properties [12,13]. Spin-filters are tunnel junctions in which the tunnel barrier height is spin-dependent because the bottom level of the conduction band in the ferromagnetic barrier material is spin-split by exchange. This allows to efficiently filter the tunneling electrons according to their spin, in other words to create a highly spin-polarized current and thus observe a large tunnel magnetoresistance effect if one of the electrodes is also ferromagnetic. Additionally, the dependence of the tunnel current on the electric polarization in the ferroelectric barrier produces an electroresistance effect. A tunnel junction with a multiferroic barrier therefore gives rise to both magnetoresistance and electroresistance effects, resulting in a four-resistance-state system.

The BiMnO$_3$ (BMO) perovskite has well-established ferromagnetic properties with a magnetic Curie temperature $T_{CM}$=105K [14]. Its ferroelectric properties are less known and discrepancies still exist on the value of its ferroelectric Curie point ($T_{CE}$=450K [15], 770K [16]). Remarkably, a magnetocapacitance effect related to multiferroicity has been observed in BMO ceramics [16]. The stabilization of the pure BMO phase is difficult to achieve either in bulk or in thin films due to Bi volatility. However, it becomes easier by partial substitution of Bi by the isovalent La, with only little influence on the physical properties at low La content [17]. We have grown LBMO films onto SrTiO$_3$(001) substrates and manganite La$_{2/3}$Sr$_{1/3}$MnO$_3$ (LSMO) buffers [18]. The LSMO layer can be used as a bottom metallic electrode for ferroelectric characterization and as a half-metallic ferromagnetic electrode in spin filters. As shown in figures 1a and 1b, LBMO films epitaxially grown on SrTiO$_3$ substrates [18] display a ferromagnetic behaviour with $T_{CM} \approx 90K$, i.e. close to the Curie temperature of bulk BMO. The saturation magnetization is lower than that of bulk BMO, as often observed in BMO and LBMO films [19,20], and likely due to Bi vacancies.

The ferroelectric nature of the LBMO films was characterized using piezoresponse force microscopy (PFM) experiments. Figure 1c shows the results for a 30 nm LBMO film. The left inset is a PFM image collected after writing negatively or positively polarized stripes in the LBMO film. A clear contrast between up and down ferroelectric domains is observed. More quantitatively, the piezoelectric phase vs electric field hysteresis cycle shown in figure 1c indicates that two remnant electric polarization states are stable in the film. The electric field dependence of the piezoelectric coefficient $d_{33}$ can be constructed by using the dependences of the phase and amplitude, (see right inset of figure 1c). The hysteresis cycle is not square as in thick ferroelectric capacitors [21], which is possibly due to the small LBMO thickness, but it undoubtedly confirms the ferroelectric character. Figure 1d shows that switchable polarization states, stable over several hours, can also be observed for a LBMO film with a thickness of only 2 nm (5 unit cells). This demonstrates the ferroelectric nature of LBMO films as thin as 2 nm. Our LBMO films are thus both ferromagnetic and ferroelectric, i.e. multiferroic.

To exploit the multiferroic character of LBMO in spintronics, we have integrated such 2 nm thick ferromagnetic-ferroelectric LBMO films as barriers in tunnel junctions using a LSMO film (30 nm) as the bottom electrode and a gold layer for the top one (see lithography details, see reference [22] and Methods). It has already been observed [11,23] that such junctions exhibit TMR effects due to spin filtering, that is, shortly, two different resistance states for respectively the parallel (p) and antiparallel (ap) orientations of the magnetization in the LBMO barrier and LSMO electrode (see figure 3a). In our samples, we recorded TMR curves at low voltage after having saturated the ferroelectric polarization of the barrier in one or the other direction by applying a large voltage (+2V or -2V) across the junction. On the TMR curves recorded at 10 mV shown in Figure 2, we can identify 4 resistance levels corresponding approximately to the 4 states represented in the right of the figure: ap magnetic configuration and a positive polarization (1), p configuration and a positive polarization (2), ap configuration and a negative polarization (3) and p configuration and a negative polarization (4).

We will now consider the possible mechanisms for the influence of the ferroelectric polarization on the tunneling current before discussing our experimental results in more detail. The influence of the electrical polarization on tunneling has been investigated from both experimental [24] and theoretical points of view [12,13]. The most intuitive mechanism leading to a modulation of the tunnel current by the polarization of the barrier is the variation of barrier thickness due to the converse piezoelectric effect. This mechanism gives rise to asymmetrical I(V) curves with a shift of the conductance minimum to a non-zero voltage [13]. From the low $d_{33}$ value estimated by PFM (see figure 1c) we expect a shift of only 3.6 mV, hardly detectable with our setup. A second mechanism for the influence of the ferroelectric polarization is related to the charge screening at the electrode-barrier interfaces and the difference of its spatial extension at both sides of the barrier. This screening controls the depolarizing field across the junction and therefore the profile of the barrier potential seen by the tunnelling electrons [12] (see figure 3b). By using electrodes with different screening lengths, the average barrier height is different for different orientations of the polarization. In the experiments of figure 2, the application of a large voltage (2V) across the junction induces a remnant ferroelectric polarization in one or the other direction, which leads to an offset of ~30kΩ between the two TMR curves recorded at a constant low voltage.

We have also recorded I(V) curves at a fixed magnetic field to look at the influence of the electric polarization induced during the voltage cycle. In figure 4a, we show I(V) and conductance curves obtained at 6 kOe by cycling the bias voltage between + and -2V (only the variation between ± 0.6V is represented in figure 4a but the I(V) curve up to 2V is shown in the inset). A noticeable hysteresis is reproducibly observed: the tunneling current is smaller (larger) when the voltage is swept from +2V to -2V (from -2V to +2V). The minimum of the conductance curves is not shifted along the voltage axis. This absence of shift confirms that the polarization-induced variation of the barrier thickness has no significant effect, as expected, and we thus conclude that the major part of the electroresistance is due to charge screening effects.

For a quantitative interpretation, we have fitted the G(V) curves of figure 4b with the Simmons model of tunneling [25], between 0.2 and 0.6V (that is out of the range in which inelastic effects induce the so-called zero-bias anomaly in magnetic tunnel junctions [26]). Our G(V) data for increasing (V<0→V>0) and decreasing (V>0→V<0) voltage sweep directions can be fitted with the same barrier thickness and two different average barrier heights: $\Phi_- = \Phi(V<0\rightarrow V>0) = 0.77$ eV and $\Phi_+ = \Phi(V>0\rightarrow V<0) = 0.80$ eV, i.e. $\Delta\Phi_{sweep} = \Phi_+ - \Phi_- = 0.03$ eV (see Fig.4b). Within the model of Zhuravlev [12], such a value of $\Delta\Phi_{sweep}$ corresponds to a LBMO polarization P=900nC/cm$^2$, taking into account screening lengths of 1 nm and 0.07 Å for LSMO and Au electrodes, respectively and a relative dielectric constant for LBMO $\varepsilon_R$=25 [8]. There are only a few data in the literature on ferroelectric properties of LBMO and BMO. Moreira dos Santos el al [15] found a polarization of 150 nC/cm² for polycrystalline BiMnO$_3$. Polarization measurements in polycrystalline samples are often difficult due to high leakage currents, which can lead to an underestimation of P. For instance, P in bulk BiFeO$_3$ is 6 µC/cm² [27] while values beyond 50 µC/cm² have been found in high-quality films [28]. A polarization of 900 nC/cm² for our LBMO film is thus reasonable.

In figure 4c, we present the variation of the electroresistance (ER) effect, i.e. the normalized difference between the I(V) curves at increasing and decreasing voltage ER=[I(V<0→V>0)- I(V>0→V<0)] / I(V>0→V>0), as a function of the bias voltage. In our experiments, the amplitude of ER increases upon increasing the maximum applied electric field (not shown), which suggests that this field is lower than the field necessary to fully polarize the LBMO barrier. On the experimental curve of fig 4c, we can distinguish a low-voltage regime were the ER is roughly constant, and a symmetric high-voltage regime where the ER decreases to zero. The inflexion points between these two regimes could either correspond to the reversal of the polarization of the ferroelectric barrier (and thus reflect the ferroelectric coercive field of LBMO), or correspond to overcoming the tunnel barrier height (and thus reflect a transition from direct tunneling to a Fowler-Nordheim regime [29]). A detailed analysis indicates that the latter mechanism is the most likely. Indeed, the position of the inflexion points match the barrier heights determined by fitting the G(V) curves (~0.8 eV, see figure 4b). Furthermore, simulations of the ER curves using Zhuravlev's model [12] and ferroelectric measurements of Moreira dos Santos el al [15] (that we reproduce in the left inset of figure 4c, normalized to a maximum polarization of 900 nC/cm²) allow to reproduce the data fairly well (see right insert of Fig.4c). These arguments strongly support our interpretation of hysteretic I(V) curves and the ER effect based on the ferroelectric character of the LBMO barrier.

Our work thus demonstrates, for the first time, that multiferroics materials are of great interest to provide additional degrees of freedom in spintronics systems. Taking advantage of both the ferroelectric and ferromagnetic characters of LBMO thin films we have obtained four different resistance states corresponding to positive or negative orientations of the ferroelectric polarization and to the parallel or antiparallel magnetizations configuration of the barrier and LSMO counter-electrode. Our work should stimulate more research on ultrathin films of multiferroic materials and on logic devices exploiting the combination of ferroelectricity and magnetism.

**Methods**

The films and heterostructures involved in this study were grown by pulsed laser deposition on SrTiO$_3$(001) single-crystalline substrates using a KrF laser ($\lambda$=248 nm) with a fluence of 2 J/cm$^{-2}$ and a repetition rate of 2 Hz. For the LBMO layers the oxygen partial pressure was set to 0.1 mbar and the substrate temperature to 625°C. For LSMO, the pressure was 0.2 mbar and the temperature 700°C. After growth, the samples were cooled down to room temperature in 1 bar of pure oxygen. Tunnel junctions were defined by spinning a thin (~30 nm) photoresist layer on the LSMO/LBMO or LSMO/STO/LBMO bilayers and indenting it with a conductive tip atomic force microscope (AFM) while monitoring the LSMO-tip resistance in real-time (see reference 22 for details). The indents where subsequently filled with ~100 nm of Au. Magnetization was measured in a Quantum Design superconducting quantum interference device (SQUID) with the magnetic field applied in the plane. Piezoresponse force microscopy (PFM) was performed with a Digital Instruments Nanoscope IV AFM with CrPt-coated conducting tips. PFM signals were extracted through the AFM signal access module to a SR830 lock-in amplifier. The phase and amplitude of the signal were collected simultaneously. The AC modulation voltage applied between the tip and the bottom electrode for reading had a typical frequency of 4 kHz and a peak-to-peak amplitude of 0.5 V. The junction transport properties were measured using a Keithley 6514 electrometer and a Keithley K230 voltage source or a Keithley 2400 multimeter, in an Oxford Instruments cryostat (3-300K) equipped with an electromagnet (0-6 kOe).

[30]    This study was partially supported by the Picasso-France-Spain program, the CICYT of the Spanish Government Project No. MAT2002-04551-C03, FEDER, the E.U. Marie-Curie mobility program, the European Science Foundation THIOX network and the E.U. STREP "Nanotemplates" Contract No. NMPA4-2004-505955.

Figure Captions

Figure 1    Temperature (a) and field (b) dependence of the magnetization of a 30 nm LBMO film. (c) PFM measurements on a 30 nm LBMO films grown on a LSMO electrode. The main panel displays the variation of the piezoresponse phase with the applied voltage. The bottom inset is the variation of the deformation coefficient $d_{33}$ with voltage and the top inset a PFM phase image of the film after applying a positive or negative writing voltage along stripes. (d) PFM phase image of a 2 nm LBMO film grown on a LSMO electrode, after writing first four voltage stripes and then two 1 µm² squares at opposite voltage.

Figure 2    Tunnel magnetoresistance curves at 4K at $V_{DC}$=10 mV in a LSMO/MBMO(2nm)/Au junction, after applying a voltage of +2V (black symbols) and -2V (red symbols). The combination of the electroresistance effect and the tunnel magnetoresistance produces a four-resistance state system. The sketches on the right indicate the magnetic (white arrows) and electric (black and red arrows) configuration.

Figure 3    (a) Sketch of the spin-dependent tunneling for parallel (left) or antiparallel (right) configurations of the LBMO and LSMO magnetizations, considering a half-metallic LSMO electrode with only spin-up states at the Fermi level $E_F$. For simplicity we assume a non-ferroelectric LBMO barrier. $\Phi_0$ is the barrier in the absence of ferromagnetism and $\Delta\Phi_{ex}$ is the exchange splitting. Spin-up is represented in red and spin-down in blue. In the parallel case, spin-up electrons tunneling from LSMO experience a small barrier height ($\Phi_0 - \Delta\Phi_{ex}/2$), which results in a large current and a low resistance. In the antiparallel case, these electrons tunnel through a larger barrier height ($\Phi_0 + \Delta\Phi_{ex}/2$), which results in a low current and a large resistance. (b) Sketch of the potential profile seen by the tunneling electrons for the two directions of the barrier electric polarization (assuming a non-magnetic barrier). $\Phi_0$ is the barrier in the absence of polarization. $\Phi_+$ and $\Phi_-$ are the average barrier heights seen by the carriers when P points towards LSMO and Au, respectively .

Figure 4    Bias voltage dependence of the current (a) and conductance (b) of a LSMO/LBMO(2nm)/Au tunnel junction, for two different bias sweep directions (negative to positive: red; positive to negative: black). Solid lines in (b) are fits to the data using Simmons model (reference [25]) beyond the zero-bias anomaly regime. Inset: I(V) curves up to the maximum applied bias voltage of 2V. (c) Electric field dependence of the electroresistance measured on a LSMO/LBMO(2nm)/Au junction. Dotted lines represent the average tunnel barrier height and separate the low-bias and the high-bias regimes (see text). Left inset: P-E loop for a $BiMnO_3$ sample, adapted from reference [15]. Right inset: simulated electroresistance vs bias voltage using the model of Zhuravlev et al [12], taking into account the screening of charges at the electrode-barrier interfaces and using data adapted from Moreira dos Santos et al [15].

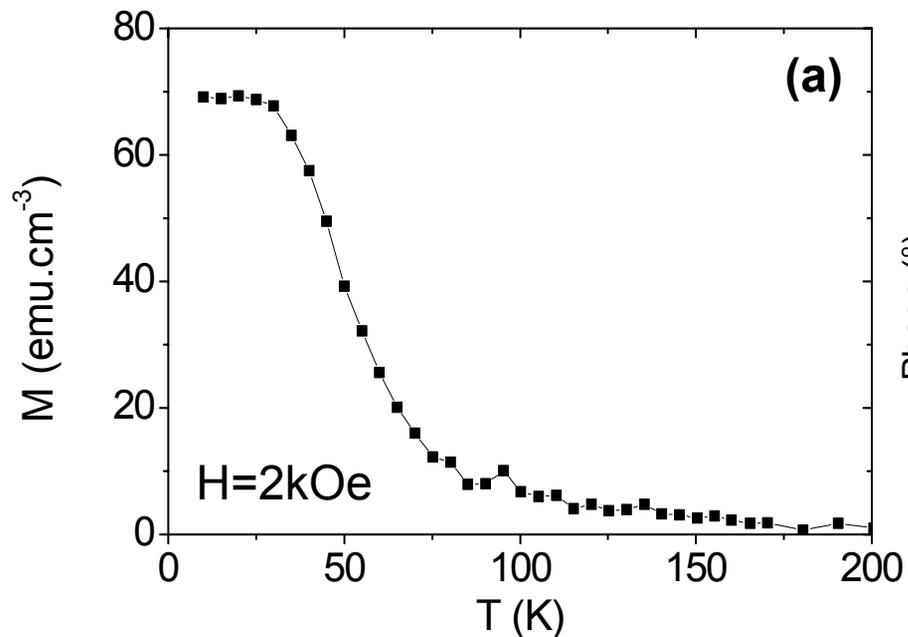
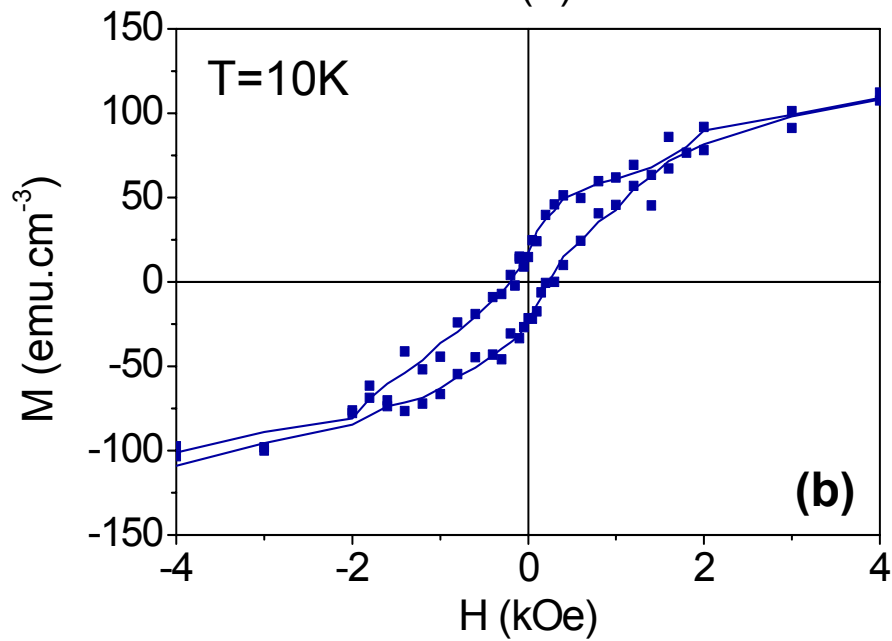
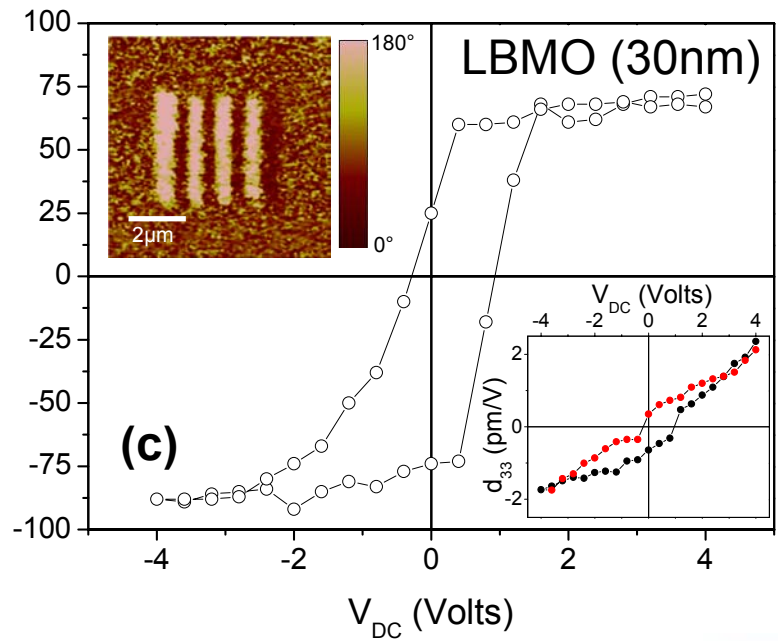
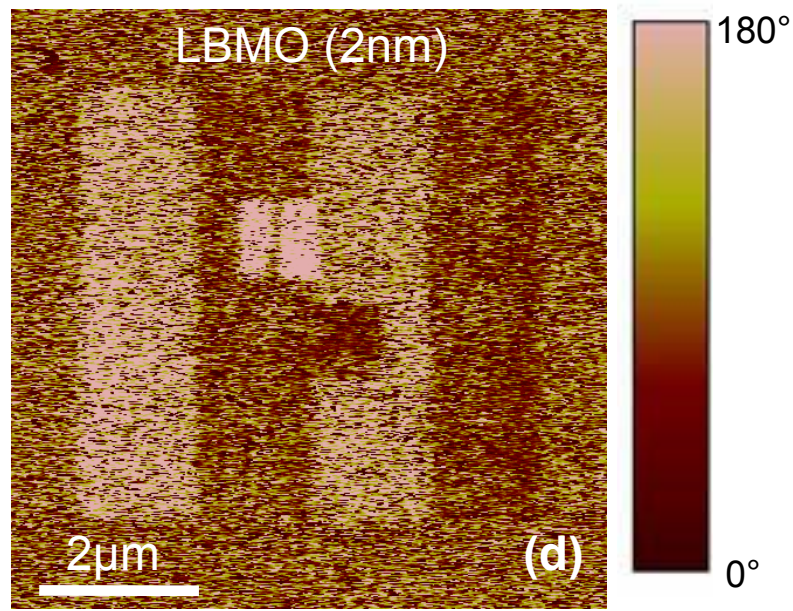

Fig. 1  Gajek et al

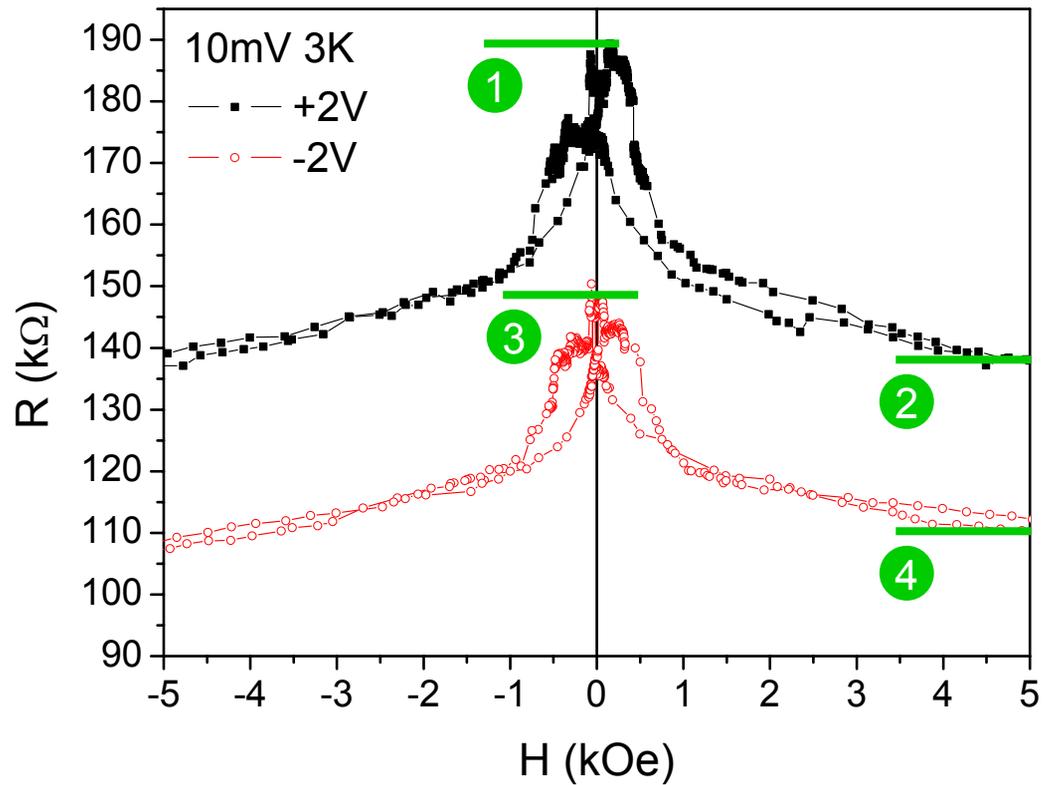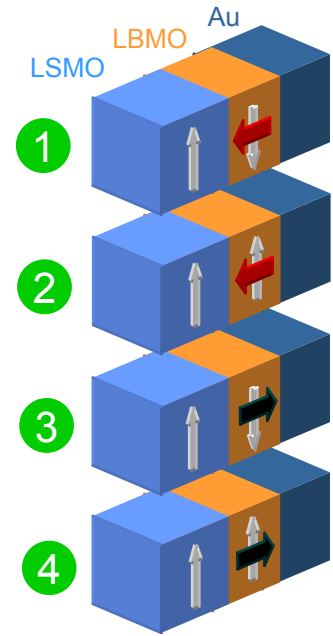

① ap configuration, positive polarization    ② p configuration, positive polarization

③ ap configuration, negative polarization    ④ p configuration, negative polarization

Fig. 2 Gajek et al

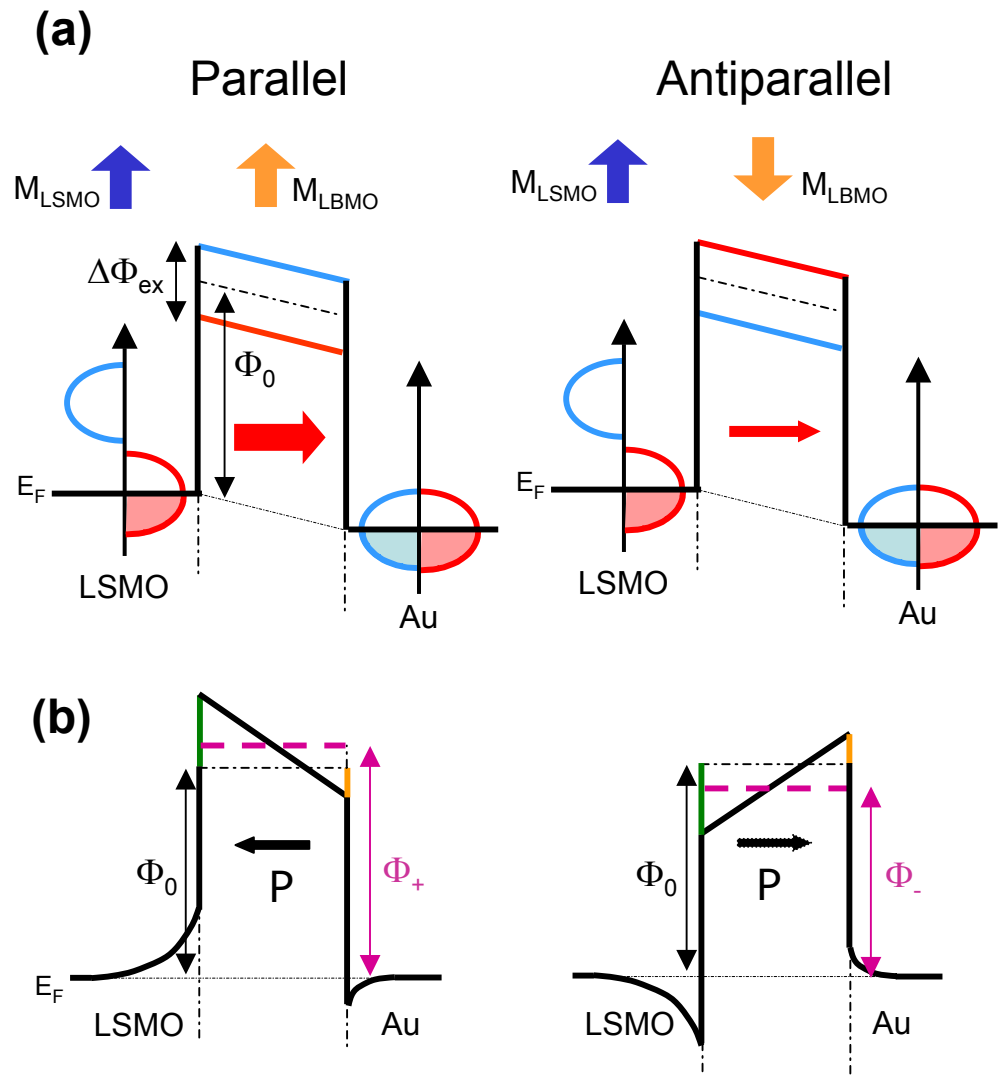

Fig. 3 Gajek et al

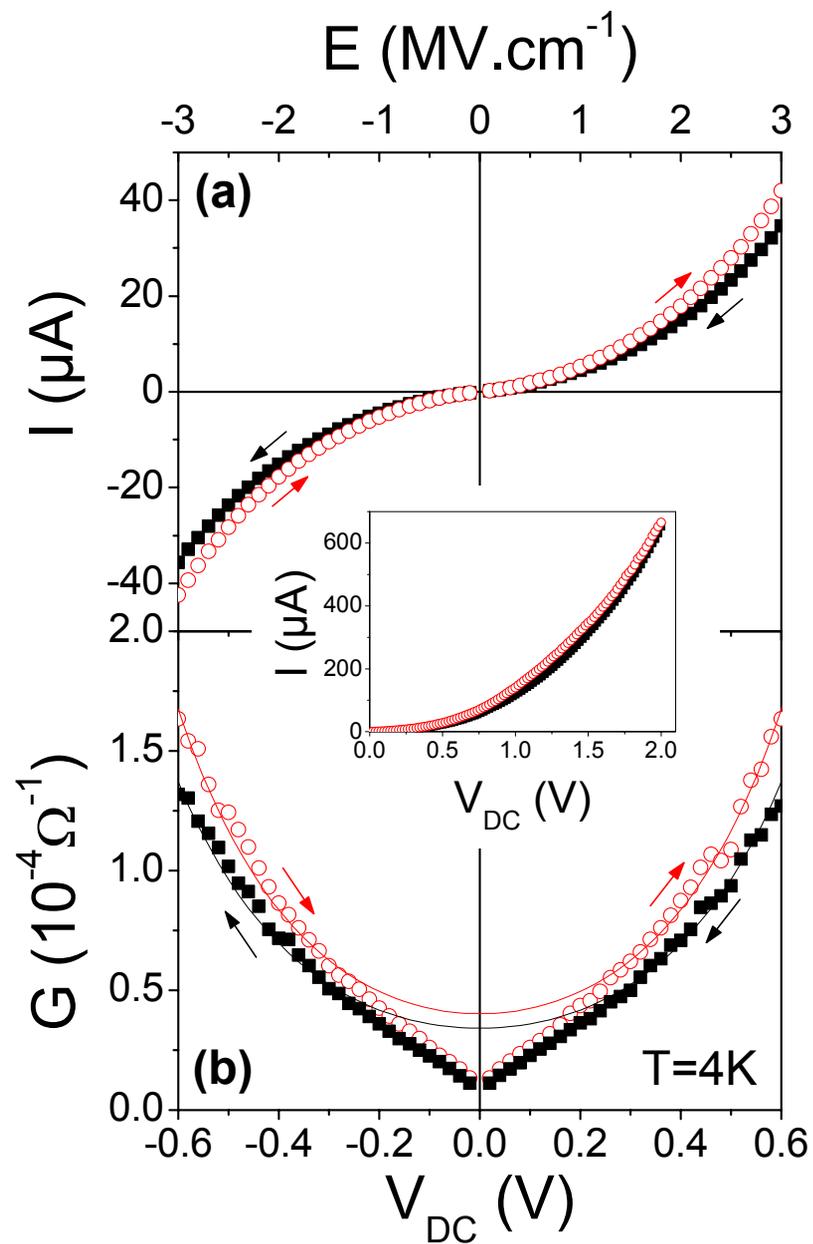
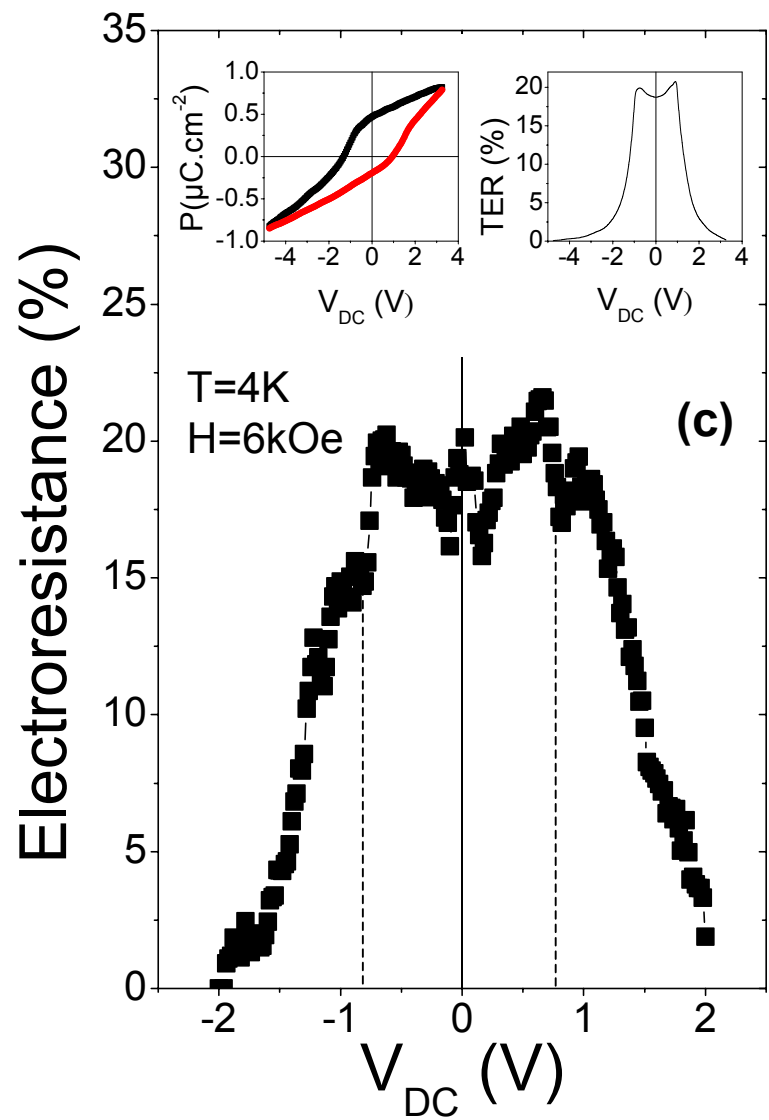

Fig. 4 Gajek et al